\documentclass{emulateapj}

\usepackage{lscape}

\slugcomment{\today}

\shorttitle{BAL Quasar 1045+352}
\shortauthors{M. Kunert-Bajraszewska et al.}

\begin{document}

\title{X-RAYS FROM A RADIO-LOUD COMPACT BROAD ABSORPTION LINE QUASAR 1045+352 AND
THE NATURE OF OUTFLOWS IN RADIO-LOUD BROAD ABSORPTION LINE QUASARS}

\author{Magdalena Kunert-Bajraszewska$^{1}$, Aneta Siemiginowska$^{2}$,
Krzysztof Katarzy\'nski$^{1}$, Agnieszka Janiuk$^{3}$}

\affil{$^{1}$ Toru\'n Centre for Astronomy, N. Copernicus University,
Gagarina 11, 87-100 Toru\'n,, Poland}
\affil{$^{2}$ Harvard Smithsonian Center for Astrophysics, 60 Garden St, 
Cambridge, MA 02138, USA}
\affil{$^{3}$ N. Copernicus Astronomical Center, Bartycka 18, 00-716
Warsaw, Poland}

\begin{abstract}

We present new results on X-ray properties of radio loud broad
absorption line (BAL) quasars and focus on broadband spectral
properties of a high ionization BAL (HiBAL) compact steep spectrum
(CSS) radio-loud quasar 1045+352.  This HiBAL quasar has a
very complex radio morphology indicating either strong interactions
between a radio jet and the surrounding interstellar medium or a
possible re-start of the jet activity.  We detected 1045+352 quasar in
a short 5~ksec {\it Chandra} ACIS-S observation. We applied
theoretical models to explain spectral energy distribution of
1045+352 and argue that non-thermal, inverse-Compton (IC)
emission from the
innermost parts of the radio jet can account for a large fraction of
the observed X-ray emission.  In our analysis we also consider a
scenario in which the observed X-ray emission from radio-loud BAL
quasars can be a sum of IC jet X-ray emission and
optically thin corona X-ray emission.
We compiled a sample of radio-loud BAL quasars that were observed in
X-rays to date and report no correlation between their X-ray and radio
luminosity. However, the radio-loud BAL quasars show a large range of
X-ray luminosities and absorption columns.
This is consistent with the results obtained earlier for radio-quiet
BAL quasars and may indicate an orientation effect in BAL quasars or
more complex dependence between X-ray emission, radio emission, and an
orientation based on the radio morphology. 

\end{abstract}


\section{Introduction}

About 10\%-30\% of quasars population show blueshifted broad absorption
lines (BALs) of the high ionization resonant lines (C\,IV 1549$\AA$ --
high-ionization BAL (HiBAL) quasars). Ten percent of them also show
absorption troughs in low ionization lines (Mg\,II 2800$\AA$ --
low-ionization BAL (LoBAL) quasars). BALs are probably caused by the
outflow of gas with velocities up to 0.2c \citep{hewett03} and are
associated with the accretion process. As it has been recently found
\citep{becker00} both radio-quiet and radio-loud quasars can have BALs.  
There are two scenarios explaining origin and nature of BAL
quasars. According to the first one, BAL regions exist in both BAL and
non-BAL quasars, and the BAL quasars are normal quasars seen along a
particular line of sight - at high inclination angles
\citep{murray95,elvis00}. The second says that BALs may be associated
with a short-lived evolutionary quasar phase characterized by a large
covering fraction wind, rather than the orientation
\citep{gregg00,gregg06}.

Theoretical models \citep{murray95,elvis00} suggest that BALs
are seen at high inclination angles, which means that the outflows 
from accretion disks are present near an equatorial
plane. However, some recent numerical work indicates
that it is also plausible to launch bipolar outflows from
the inner regions of a thin disk \citep[e.g.][]{ghosh07,prog04}.
There is a growing observational evidence indicating
the existence of polar BAL outflows \citep[e.g.][]{zhou06,ghosh08}. 
This means that there is no one simple orientation model which can explain all
the features observed in BAL quasars.

X-ray observations of radio-quiet BAL quasars
\citep{green01,gall06} indicate that they are X-ray weak,
perhaps due to a high absorption column intrinsic to the source.
Their X-ray spectra display a clear evidence of X-ray absorption,
often complex, with high intrinsic column densities
$N_{H}>10^{22}~{\rm cm^{-2}}$ \citep{gall02}. The origin of the BAL
X-ray absorber is still debated, although a broadening of the optical
absorption lines suggest the high velocity outflows as a source of
this absorption. This view was challenged by
\citet{gius08} and \citet{shen08} who concluded that X-ray selected
BALs appear less X-ray absorbed than purely optically selected
ones. This could be a geometrical effect: we observe these objects 
at smaller angles with respect to the accretion disk rotation axis, 
so missing the bulk of X-ray shielding gas and the highest outflow velocities.

Radio band observations can provide orientation information for
radio-loud quasars with BALs, so chances to study and identify the
origin of the absorbing gas are higher in the radio-loud than in the
radio-quiet quasars.  In addition radio-loud quasars are more X-ray
loud than the radio-quiet quasars
\citep{young09}.  However, X-ray observations
of radio-loud BALs are sparse and only a handful of them have been
observed in X-rays so far \citep[][and this paper]{broth05,mill06,
schaer06,wang08}.  There is also a very limited information about the
radio morphology of BAL quasars. There are only 10 of them with known
large scale radio morphology
\citep{wills99,gregg00,gregg06,broth02,zhou06}.  Most of the
radio-loud BAL quasars detected to date have very compact radio
structures similar to Gigahertz-Peaked Spectrum (GPS) and compact
steep spectrum (CSS) sources which are thought to be young
\citep{odea98}. There are only a few radio-loud compact BAL quasars
with resolved structures that were observed in radio band at one
frequency only
\citep{jiang03,liu08}. The first multifrequency radio observations 
of a very compact HiBAL quasar 1045+352 were made by \citet{kun07}.
The complex compact structure has been resolved into many
sub-components and indicates that the jet is moving in a non-uniform
way in the central regions of the host galaxy.

We observed 1045+352 BAL quasar ($z=1.604$) in X-rays using {\it
Chandra} X-ray Observatory.
We report the results of this
observation in Sec.~\ref{obs}, present and model the spectral energy distribution
(SED) of 1045+352 in Sec.~\ref{quasar} and discuss the source properties together with
the properties of all radio-loud BAL quasars observed to date in Sec.~\ref{dis}.
Finally Sec.~\ref{summary} summarizes the main conclusions.

Throughout the paper, we assume a cosmology with 
${\rm
H_0}$=71${\rm\,km\,s^{-1}\,Mpc^{-1}}$, $\Omega_{M}$=0.27,
$\Omega_{\Lambda}$=0.73.

\section{{\it Chandra} X-ray Observation}
\label{obs}

1045+352\ BAL quasar was observed with the {\it Chandra} Advanced CCD
Imaging Spectrometer \citep[ACIS-S,][]{garmire03} on 2008 Jan 20
(ObsID 9320). The source was located at the default aim-point
position on the ACIS-S backside illuminated chip S3 (Proposer's
Observatory Guide
(POG)\footnote{http://asc.harvard.edu/proposer/POG/index.html}). The
1/8 subarray CCD readout mode of one CCD only was used resulting in
0.441~sec frame readout time.  The effective exposure time for this
observation was 4658.44 sec.

We observed five counts in the assumed 1.3~arcsec radius circular source
region defined at the position of the radio source
(Table~\ref{table1}). All the detected counts are within the 0.5-7~keV
energy range.  We calculated the significance of this detection
assuming the Poisson distributions for the source and
background. Given 6 background counts in the background region ($\sim
50$ times larger than the source region) and assuming the Poisson
background we simulated the expected background count rate in the
source region. This simulation accounts for the fluctuations in the
background. We obtained the significance of observing five counts given
the background rate at 6$\times 10^{-7}$ and concluded that the
detection of the source is highly significant.

The observed five counts give a source flux within 0.5-7~keV of
$1.25^{+0.37}_{-0.58} \times 10^{-14}$~erg~cm$^{-2}$~s$^{-1}$ assuming
a power law model with $\Gamma=1.7$
and absorption at $z=0$ of $N_H=1.28^{+1.06}_{-0.7} \times
10^{22}$~cm$^{-2}$ (the errors
are 1$\sigma$ for 1 significant parameter). 
The absorption column is treated as an observed minimal value of the
absorption in this source and the calculated source flux we take into account in
SED modeling.
The absorption column was obtained in Sherpa by fitting an absorbed power law
model with fixed photon index assuming Cash statistics and it is only a
rough approximation as the true
value of the continuum shape is unknown. We derive the 3 sigma upper
limit on the intrinsic absorber at $z=1.604$ of $N_H < 9.6\times
10^{23}$~cm$^{-2}$.
The unabsorbed X-ray flux in the 0.5-7 kev band assuming
only Galactic absorption and photon index $\Gamma=1.7$ 
(e.g. fitting only the model normalization)
is $7.00\times10^{-15}$~erg~cm$^{-2}$~s$^{-1}$.

We calculated the observed hardness ratio for the source:
${\rm HR}=(H-S)/(H+S) = 3/5 = 0.6$
(where, H= hard(2-10 keV) = 4 counts; S= soft(0.5-2keV) = 1 count).
The X-ray spectrum appears to be relatively hard, although more counts are
needed to confirm this result.
Table~\ref{table1} lists all the observed X-ray and radio
parameters for this source.

\section{BAL Quasar 1045+352}
\label{quasar}

1045+352 HiBAL quasar is a compact radio-loud CSS source at redshift
z=1.604 \citep{willott02}. According to the evolution theory the CSS sources are young
AGNs evolving into large scale FRI and FRII radio sources. They
probably originate in a merger event and it is possible that in a case
of 1045+352 traces of that process are still visible. This HiBAL quasar
1045+352 has a complex radio morphology with a radio jet axis
reorientation: the NE/SW emission can be the first phase of activity,
now fading away, and the extension in the NW/SE direction is a
signature of the current active phase. Such a reorientation of the jet
axis may result from (1) a merger, (2) a jet precession, or (3)
jet-cloud interactions \citep{kun07}. Its high submillimeter emission
and a distorted radio structure indicate that the whole source resides inside
the dense medium of the host galaxy, which together with the small
(4.3\,kpc) linear radio size suggest it is a young active galactic nucleus
(AGN).

Our {\it Chandra} X-ray observation of 1045+352 gives a
source unabsorbed 2\,keV flux of
1.32$\times10^{-15}$~erg~cm$^{-2}$~s$^{-1}$~keV$^{-1}$ assuming a
power law model with $\Gamma=1.7$ and Galactic absorption. 
The estimated observed $\alpha_{ox}=1.38$ would indicate a rather normal,
unabsorbed radio-loud quasar. 
However, the optical flux is reddened by $A_{V}=1.5$ \citep{willott02}. Based
on the estimated dereddened optical flux we calculated intrinsic
optical$-$X-ray index, $\alpha_{ox}=1.88$. This is typical value
for radio-quiet quasars, although \citep{siemi2008} found
the similar values of $\alpha_{ox}$ in a sample of radio-loud GPS and
CSS quasars. Note that such high value of $\alpha_{ox}$ may also
suggest a presence of an X-ray absorber close/in the BLR region.

\subsection{Black Hole Mass Estimation}

Using the published optical spectrum and emission line data for
1045+352 \citep{willott02} we estimated its black hole mass and the
Eddington luminosity ratio. To estimate the black hole mass we used
the following prescription derived by \citet{McLure02}, which uses the
3000${\rm
\AA}$ monochromatic luminosity and the Mg\,II $\lambda$2800\,FWHM:

\begin{equation}
{\rm
\frac{M_{BH}}{M_{\bigodot}}=3.37\left(\frac{\lambda L_{3000}}{10^{44}
erg~s^{-1}}\right)^{0.47} \left(\frac{FWHM(Mg\,II)}{km~s^{-1}}\right)^{2}
}
\end{equation}

\noindent
The obtained value is ${\rm M_{BH}=1x10^8~M_{\bigodot}}$, which we then used to calculate the
Eddington luminosity:

\begin{equation}
{\rm
L_{Edd}=1.25x10^{38}\left(\frac{M_{BH}}{M_{\bigodot}}\right)~erg~s^{-1}
}
\end{equation}	

\noindent
We estimated the bolometric
luminosity for 1045+352 using the following equation:

\begin{equation}
{\rm
L_{bol}=4\pi D_{L}^{2} f (1+z)\lambda F_{\lambda},
}
\end{equation}

\noindent
where $\lambda=3000{\rm \AA}$ and f=5 is the average bolometric correction from
3000${\rm \AA}$ \citep{ganguly07}. The Eddington luminosity ratio is then:
${\rm L_{bol}/L_{Edd}=0.36}$.
It has to be noted here that the mass estimate based on the Mg\,II emission
line width (and so the value of the accretion rate) can have a large error.
We treated these values as the input parameters for an accreting corona model
of active galactic nuclei described by \citet{janiuk} which we used together
with the synchrotron self-Compton scenario to model SED of 1045+352.

\subsection{Broad-Band Emission of 1045+352: Models}

Broadband spectra of 1045+352 are characterized by the strong radio
emission, peak in the IR, and relatively strong UV-X-ray continuum. We
apply two emission models to the observed multiwavelength spectrum of 
1045+352. We start with the emission related to the jet,
because the quasar is radio-loud and the jet dominates the observed
radio emission. In the second step we calculate the contribution to
the spectrum from our accretion disk-corona model.

\subsubsection{X-ray Emission of the Jet}

Significant or even dominant part of the X-ray emission could be
produced by propagating radio jets and expanding lobes,
located at the distance a few hundred parsec from
the center of the source \citep{mill06,stawarz08}. To check this possibility 
we use a simple model that is able to reproduce the observed spectra.

We assume a spherical geometry of this part
of the jet, described by radius $R$. More complex, conical or
paraboloid geometry could better describe this region
\citep[e.g.][]{Konigl81,Ghisellini85}, however, such modeling
requires more free parameters, which are difficult to constrain from
the observations. The spherical region in our model is filled by
relativistic electrons and tangled magnetic field characterized by
strength $B.$ The electron energy distribution is assumed to be a
power-law function $N (\gamma) = K \gamma^{-n}$ for $1 < \gamma
\leqslant \gamma_{\max}$, where $\gamma$ is the Lorentz factor that
describes electron energy ($E = \gamma m_e c^2$), $K$ describes
particle density for $\gamma = 1$, and $n$ is the index of the energy
spectrum. The relativistic electrons spinning around magnetic field
lines produce synchrotron emission. Some fraction of this emission is
upscattered to higher energies by the electrons. This is well known
synchrotron self-Compton scenario (SSC). To calculate the synchrotron
emission we use the emissivity derived by \citet{Crusius86} and
\citet{Ghisellini88}. The inverse-Compton (IC) radiation is calculated
using Compton-kernel derived by \cite{Jones68}. The emission produced
in the jet comoving frame is transformed to the observer's frame using
standard formulae and assumed value of the Doppler factor
($\delta$). However, proper motions of jet's components in CSS
objects are relatively slow \citep[up to 0.3c,][]{pol03}. This
together with an estimation of the viewing angle (see Table
1) is limiting the Doppler factor to $\delta <2$ and practically
eliminating it as a free parameter.

This is in fact the simplest possible scenario that may explain jet
emission in a wide range of energies, starting from low frequency
radio observations up to gamma rays. The model requires only three free
parameters ($R, B, K$).  Moreover, the index of the particle energy
spectrum ($n$) can be constrained directly from the observations using
fundamental relationship $\alpha_{r}=(n-1)/2$, where $\alpha_r$ is the
spectral index of the observed synchrotron emission $F_{\rm syn}
\propto \nu^{-\alpha_{r}}$. The maximum energy of the electrons,
characterized by $\gamma_{\max}$ can be derived from the standard
formula that describes particle cooling due to the synchrotron and the IC emission

\begin{equation}
\gamma_{\max}(t) = \frac{1}{C t}, \; \; C = \frac{4}{3} \frac{\sigma_T c}{m_e
c^{2}} \left(U_B + U_{\rm  rad} \right),
\end{equation}

\noindent
where $U_B = B^2/(8 \pi)$ is the magnetic field energy density and $U_{\rm
rad}$ is the synchrotron radiation field energy density 
\citep[i.e.][]{Kardashev62}. 
We calculate $\gamma_{\max}$ for the time $t=R/c$, e.g. the
minimum travel time across the source. This assumption may
overestimate the maximum energy if the cooling process is dominating
the particle energy evolution. On the other hand our estimation does
not take into account any possible acceleration of the particles that
may partially compensate the cooling process.

In our first attempt to fit the observations we assume that the whole
emission observed from the radio frequencies up to IR range is
produced by a spherical knot of the jet. This assumption gives five
direct observational constraints. The first constraint is the
self-absorption frequency ($\nu_s$). Some fraction of the low frequency
synchrotron emission can be absorbed by the electrons inside the
source. This is well known electron self-absorption process that
modifies index of the synchrotron emission producing
$\alpha_{r}=5/2$. In 1045+352 the self-absorption modifies the index
of the emission below $\nu_s \simeq 10^8$ Hz. On the other hand, the
synchrotron emission of the knot must extend up to the IR range, above
$10^{11}$ Hz. This gives observational constraints for $\gamma_{max}$
that must be high enough to explain such emission. The levels of the
synchrotron and the IC emission give the next two observational
constraints. Note that the ratio between the synchrotron and the IC
emission directly indicates the ratio between $U_B$ and $U_{\rm rad}$

\begin{equation}
\frac{U_B}{U_{\rm rad}} \simeq \frac{F_{\rm peak, syn}}{F_{\rm peak, IC}},
\end{equation}

\noindent
where $F_{\rm peak} = \max[\nu F(\nu)]$. In the case of 1045+352
$F_{\rm peak, \; syn} > F_{peak, \; IC}$ that shows dominance of the
synchrotron cooling over the IC looses. This basically means that the
value of $\gamma_{\rm max}$ is controlled by the magnetic field
strength. Finally, the slope of the synchrotron emission gives the
fifth observational constraint. We estimated the value of this
parameter to be $\alpha_{r}=0.8$, that gives $n=2.6$. The $\alpha_{r}$
value has been determined using all available radio points and it is
slightly different from that presented in Table~\ref{table1}. Taking
into account all the above constraints we obtained satisfactory fit
(Fig.~\ref{corona}a) using $\delta = 1.3$, $B=8 \times 10^{-4}$~G, $K
= 2.5$ cm$^{-3}$ and $R=1.1 \times 10^{21}$ cm (356.5\,pc). Moreover,
we calculated values of $\gamma_{\max} = 2.68 \times 10^{4}$, $U_{\rm
rad}/U_{B} = 0.23$ and $E_e/E_B = 1.34 \times 10^2$, where $E_e$ is
the total energy of the electrons and $E_B$ is the total energy
accumulated in the magnetic field. The last result shows no
equipartition between the particle energy and the magnetic field
energy. The equipartition would require a few times larger value of
the magnetic field. However, such strong magnetic field could cause
efficient synchrotron cooling, reducing significantly maximum energy
of the particles. On the other hand the equipartition 
does not have to be reached in a
compact region of the jet, located at a relatively small distance from
the center, where we still may expect the particle acceleration. 
Note that it is possible to reduce a
disagreement between $E_e$ and $E_B$ by considering for example twice as large
source and reducing a few times the particle density. This will also
significantly decrease the level of the IC emission because this emission
is proportional to the square of the particle density. Therefore, the
source in equipartition will not produce efficiently the X-ray
emission. However, a bigger source needs more time to be created 
and therefore the particles have more time to be cooled. This reduces
significantly $\gamma_{\rm max}$ and makes impossible to explain 
observed level of the emission above $10^{11}$ Hz. More extended 
synchrotron source is also optically thin at lower frequencies. The
self-absorption break ($\nu_s$) in such spectrum appears below $10^8$ Hz 
which does not agrees with the observations. In other words the bigger source
has synchrotron spectrum $"$shifted$"$ toward lower frequencies.

The radius of the jet knot derived in our first fit is relatively
large (356.5 pc), a few times larger that the size of the smallest
jet component tightly constrained by 
the high resolution radio maps 
\citep[60-70\,pc;][]{kun07}. Therefore in
our second fit we assume relatively small radius to be in agreement
with the radio observations. The reduced volume of the source must be
compensated by increased particle density and/or increased magnetic
field strength. However, $B$ cannot be increased significantly because
this parameter has direct impact on the maximum energy of the
particles. The density can be increased significantly but this will
make the source optically thick for the low frequency radio emission,
shifting $\nu_s$ toward higher frequencies. On the other hand the low
frequency radio emission can be produced by more extended structures
of the jet. Unfortunately we have only the spectrum of the total
emission of 1045+352 without information about the contribution of
separated jet components to the total emission. The best fit
(Fig.~\ref{corona}b) in this particular case was obtained for $\delta
= 1.5$, $B=3.0 \times 10^{-3}$ G, $K = 19$ cm$^{-3}$ and $R=2.1 \times
10^{20}$ cm (68\,pc). We also calculated values of $\gamma_{\max} =
1.0 \times 10^{4}$, $U_{\rm rad}/U_{B} = 0.21$ and $E_e/E_B =
72.0$. Note that the disproportion between $E_e$ and $E_B$ was
slightly reduced. However, further reduction of the source size is not
possible.  More compact and therefore more dense source would produce
efficient X-ray emission, significantly above the observed level. 

We assume that well visible excess of the emission in the
IR range is produced by dust in the center of 1045+352.
Therefore we approximate this emission using a greybody
spectrum that differs from a blackbody ($B_{bb}(\nu)$)
distribution by a simple factor

\begin{equation}
B_{gb}(\nu) = (1-e^{-\tau(\nu)}) B_{bb}(\nu),
\end{equation}

\noindent
where

\begin{equation}
\tau_{\nu}=\left(\frac{\nu}{\nu_{o}}\right)^{\beta}=\left(\frac{\lambda_{o}}{\lambda}\right)^{\beta}
\end{equation}

\noindent
$\nu_{o}$ is a turn-over frequency (or $\lambda_{o}$, the
turn-over wavelength), the frequency at which the source becomes
optically thin, and $\beta$ is the emissivity index of the dust
grains. We use in our calculations $\beta=2$
that changes index of the power-law part in the spectrum from
$\nu^2$ in classical blackbody spectrum to $\nu^4$ in the
greybody distribution that better describes thermal emission
of the dust. This is in fact the only one difference between
the blackbody and greybody spectrum in our approach.

To calculate possible influence of the greybody emission for
the IC scattering inside the jet component that
produces the X-ray emission, we use approximation proposed by
\citet{ino}. The only one difference between this
approximation and our calculations is that we use greybody
distribution instead of the blackbody one. 
According to this approximation the emission is
characterized by luminosity of the center ($L_{\rm nuc}$) and
the temperature of the dust ($T$). Moreover the intensity of
the emission in the comoving frame of the source depends on
the distance to the center $D$.

The best spectrum of the observed IR emission was obtained
for $\nu_{\rm peak} L_{\rm nuc} (\nu_{\rm peak}) = 8 \times 10^{45}
~erg~s^{-1}$ and the dust temperature $T=50$\,K. Moreover, we found that
the inverse-Compton scattering of the photons produced by the
dust becomes negligible for $D > 2 \times 10^{21}$ cm. Note that
the radius of the source is $R=2.1 \times 10^{20}$ cm thus
such relatively large source should not be placed to close
to the center and therefore IC scattering of
the dust photons can be neglected.

The optical data (starlight from the host galaxy) are taken
from the SDSS and fitted with the blackbody curve with a temperature
$T=10^{4}$\,K and $\nu_{\rm peak} L_{\rm nuc} (\nu_{\rm peak}) = 10^{45}
~erg~s^{-1}$. The IC scattering of this additional radiation field appears
also negligible for $D > 2 \times 10^{21}$ cm.

The two cases described above and illustrated in Fig.~\ref{corona}a,b
show that the high energy emission of a radio-loud BAL quasar can be
explained by a simple SSC model. The observed synchrotron emission
constrains the free parameters of the model well. Therefore the calculated
level of the IC emission is also well constrained and appears
to be in a good agreement with the observations. In other words, 
it is difficult to explain the observed synchrotron spectrum
of the synchrotron emission and simultaneously to avoid significant emission
in the X-ray range due to the SSC scattering.
The possibility that the X-ray emission in young GPS sources is produced by
a jet was recently discussed \citep{tengstrand}, but there was no
straightforward answer so far.

\subsubsection{X-ray Emission of the Corona}

We use an accreting corona model derived by \citet{janiuk} to model
the X-ray emission in 1045+352 (see also other applications of this
model: \citet{janiuk99,bechtold}).
Following this model, we assume that the (stationary) accretion is
ultimately responsible for the broadband emission, and the flow
consists of two media: relatively cold, optically thick disk, and hot,
optically thin corona.  This model has only three basic parameters: an
accretion rate, $\dot{M}$, a black hole mass, M, and the viscosity
parameter, $\alpha$.  The fraction of the energy released in the
corona is not a free parameter, but is calculated as a function of
radius from the global parameters. All measurable quantities, like the
ratio of the disk emission to the coronal emission, the spectral
slopes and the extension of the spectrum into gamma-ray band result
from the model, including the trends for the change of these
quantities with accretion rate. In the case of limited spectral
information a small number of model parameters is a big advantage over
the other more complex disk-corona models \citep[e.g.][]{sobol1,sobol2}.

We assume that the hot optically thin corona is a two-temperature
medium, i.e. the ion temperature is higher than the electron
temperature \citep{shapiro}. The loss of gravitational energy by
accreting coronal gas is transported directly to ions, while the
Coulomb coupling transfers this energy to electrons and finally
electrons cool down by the IC process. The disk emission
provides the source of soft photons for Comptonization, and the
amplification factor is defined by the Compton parameter $y$
\citep{witt}.

We assumed a non-rotating black hole so the inner disk radius is located at 
3$R_{Schw}$ and the Eddington accretion rate is defined using the efficiency of 
accretion equal to 1/16:

\begin{equation}
\dot m = {\dot M \over \dot M_{\rm Edd}} = {\dot M \over 3.52 M_{8}}
\end{equation}

\noindent
where $M_{8}=M/10^{8} M_{\odot}$ is the black hole mass, and $\dot M$ is
given in the units of solar masses per year. The non-rotating black hole
model is taken here for simplicity, although in the case of the radio-loud 
QSOs the rotating black hole is usually considered.
The spin of the black hole would result in shifting of the inner radius of
the disk-corona system closer to the black hole. However, the coronal
emission is dominated by the outermost parts of the corona, and the X-ray
spectrum would not change due to the black hole rotation.
Therefore the outcome of the modeling is practically unaffected by the value
of the black hole spin.

Overall, the model spectra do reproduce the observed AGN spectra,
as most of the emission 
is released from the accretion disk in the UV and soft X-ray bands and some 
fraction of the total luminosity is emitted in hard X-rays because of 
Comptonization. These two components, soft and hard, are modeled simultaneously
due to the radiative coupling between the disk and corona. 
The initial input parameters to the model ($M_{BH}=
1\times10^8 M_{\odot}$ and $\dot m =0.36$) were estimated from the
spectrum of 1045+352 \citep{willott02}.
The assumed viscosity parameter, $\alpha$, was either 0.1 or
0.03. The disk emission was also corrected for the galactic starlight,
by means of the thermal emission which peaks at $\log \nu = 14.5$
[Hz]. However, for these initial parameters we
did not obtain a satisfactory fit, because the disk luminosity was too
high to reproduce correctly the optical emission detected in
1045+352. Also, the coronal emission was too high in X-rays and
exceeded the {\it Chandra} limits.

By decreasing the accretion rate 
we obtained a lower flux emitted from the accretion disk, but it didn't
help us to fit the optical data. 
Note that the ratio of the luminosity emitted in the corona with
respect to the disk emission is higher for a smaller accretion rate.
This is because the Compton cooling
of the corona is less efficient due to a lower soft photon flux from
the disk, so the electron temperature in the corona can be higher.

The black hole mass does not influence the shape of the spectrum, but
a smaller black hole mass helped to shift the disk peak emission
toward somewhat higher energies and better fit the data.
To match the optical data we needed to lower the black hole mass to
$2\times10^7 M{\odot}$. This increased the accretion rate, which we had to
again decrease and finally we got the best fit with the initial input
parameter $\dot m=0.36$.

The spectral shape is affected by both viscosity $\alpha$ and accretion rate $\dot m$. 
For small viscosity, 
the relation is roughly monotonic, i.e. the larger $\dot m$, the softer 
the spectrum. This is because for large $\dot m$ the radial extension of 
the corona is larger. The hard X-rays are generated mainly at large radii 
from the black hole, because at the outer radius of the corona the
fraction of the energy dissipated in the corona is $f_{cor} (r_{out})\approx 1.0$, 
while at smaller radii $f_{cor}(r)$ is much lower.
For large $r_{out}$ the gravitational energy available for dissipation is
smaller, so there is less flux in hard X-rays. 
Therefore, with the same electron temperature in the corona at 
$r_{out}$, we have a much more profound disk emission, that forms the big 
blue bump and the spectral index is steeper. For a large $\alpha$, this 
dependence is not always monotonic. This is because the radial extension of
the corona only very weakly depends on viscosity, while the electron
temperature decreases with $\alpha$.

\begin{figure*}
\centering
\includegraphics[width=12.3cm, height=9cm]{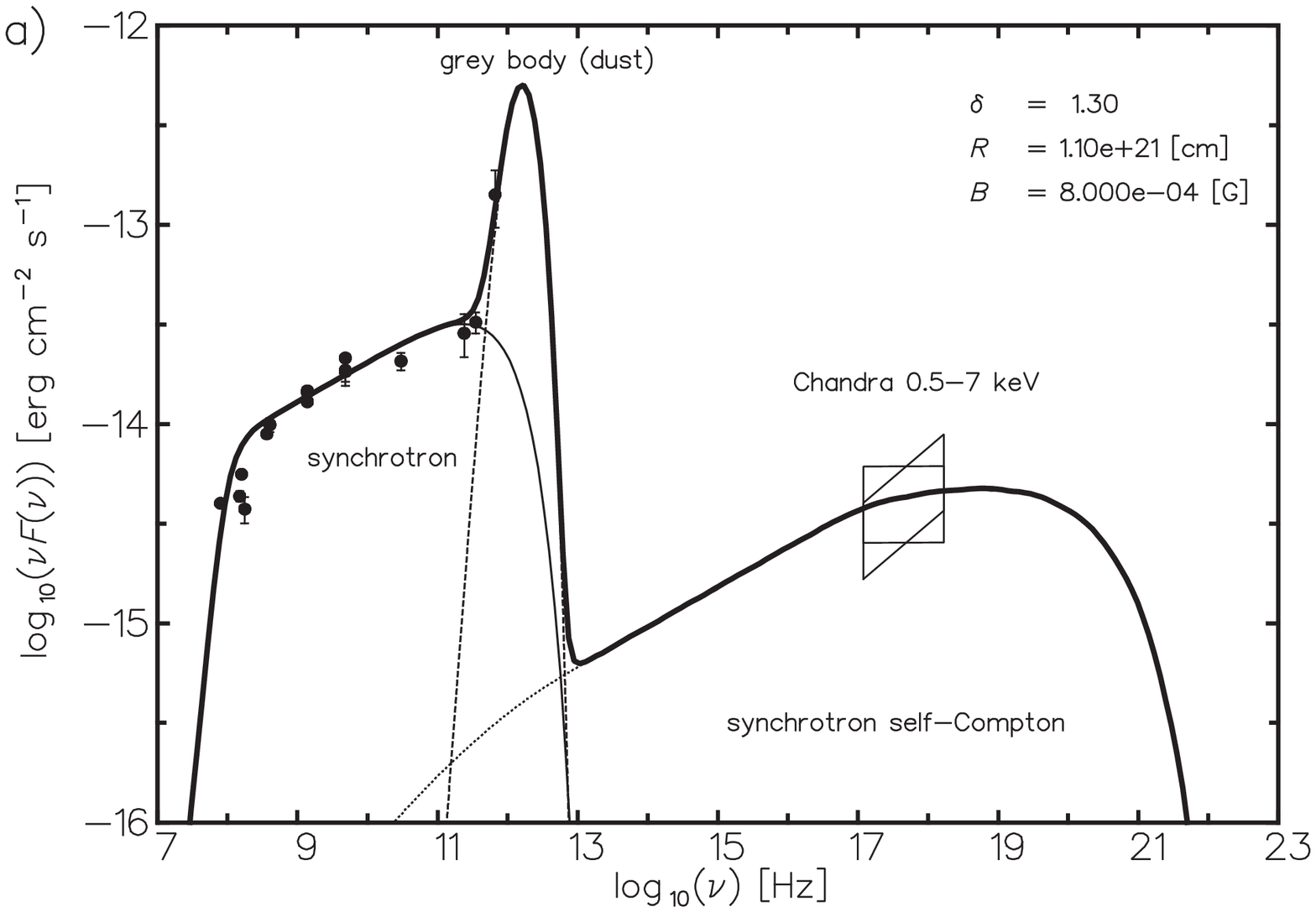}
\includegraphics[width=12cm, height=9cm]{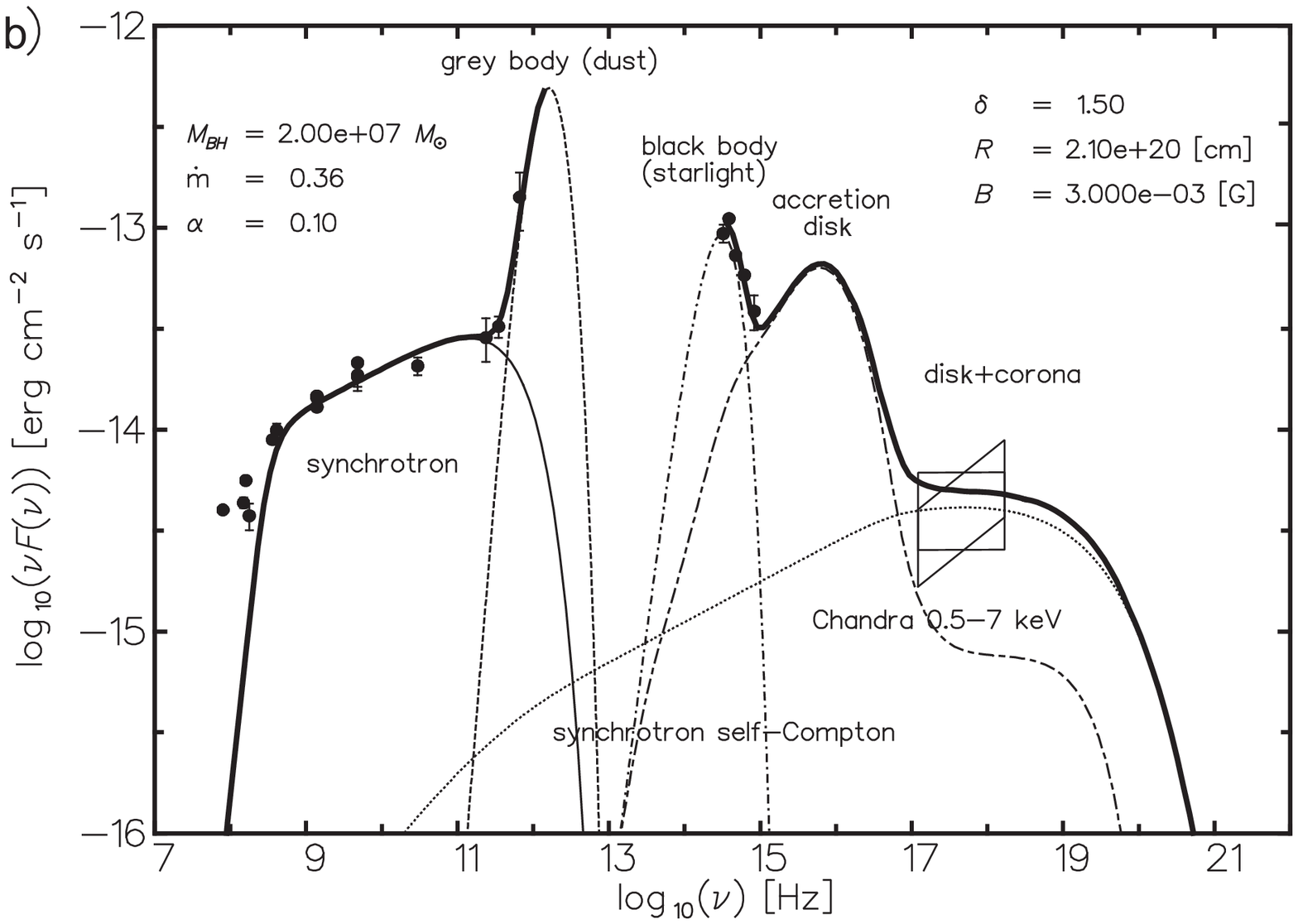}
\caption{{\footnotesize
Spectral Energy Distribution of 1045+352
based on the data gathered from the literature and SDSS.
Our {\it Chandra} observations are presented as a box: width means
0.5-7\,keV range, hight is a flux uncertainty, and
the box is drawn for $\Gamma$=1.7 and $\Gamma$=2.
The curves show the
modeled spectral components: synchrotron emission (solid line), and
corresponding SSC emission (dotted line), thermal dust emission (dashed
line); thermal starlight (dash-dotted line), and comptonized thermal
emission from the accretion disk and corona (short-long dashed line).
The bold solid line is the sum of the shown spectral components. Since there
is a lack of the infrared data the plot is not drawn in that regime.
The indicated parameters are: $\delta$ - Doppler factor, R - source
size, B - magnetic flux, $M_{BH}$ - mass of the black hole, $\dot{m}$ -
accretion rate, and $\alpha$ - viscosity parameter.
The values of the fluxes are K-corrected.
}}
\label{corona}
\end{figure*}

Because of the above complex dependencies, we had to change also the
viscosity parameter to fit the data best (Fig. \ref{corona}b).  For a
too small viscosity, the corona is rather compact and hot, so the hard
X-ray flux was too large. Also, the viscosity cannot be too small:
for $\alpha \le 0.01$ the corona became
too small and we could not determine the corona structure. We have
applied two values of viscosity parameter to our model: $\alpha =
0.03$ and $\alpha=0.1$.  We found, that for the black hole mass and
accretion rate of $M_{BH}=2\times 10^{7} M_{\bigodot}$ and $\dot
m=0.36$, both viscosity coefficients give the X-ray flux within the
{\it Chandra} limit, however a better fit is obtained for the larger value
of $\alpha=0.1$.  We note that both these values are plausible for AGN
disks. A moderately small value of $\alpha\sim 0.02-0.04$ is in
agreement with the results of the MHD simulations
\citep{turner, hirose}.  On the other hand, the observational
constraints for the $\alpha$ parameter give even larger values, of
the order of 0.1-0.4
\citep{king}, therefore such values are also plausible.

The best model spectrum for 1045+352 is presented in
Fig.~\ref{corona}b and with the best parameters:
$M_{BH}=2\times10^7\,M_{\bigodot}$, $\dot{m}=0.36$ and $\alpha=0.10$.
showing that the observed X-ray emission could be due to both the disk
corona and the relativistic jet. However the relativistic jet
emission dominates over the corona emission in X-rays. As we discussed
in the previous paragraph the jet SSC emission 
is strong and well constrained by the observed synchrotron spectrum.
On the other hand 
the optical and X-ray data can be affected by the extinction in the
source nuclear region and the intrinsic value of the emission of the
accretion disk and corona can be higher.
Deeper X-ray observations could allow us to estimate the X-ray emission
with more accuracy and help us to differentiate between model possibilities.   

As we show in the discussion
in Sec.~\ref{dis} the X-ray emission of 1045+352 is very weak
comparing to the other radio-loud BAL quasars suggesting a strong
absorption. If so the X-ray emission we observe can be mostly due to
X-ray emission from the relativistic jet, while the X-ray emission
from the corona is absorbed in a large part.

\section{Discussion}
\label{dis}

\begin{figure}
\centering
\includegraphics[width=\columnwidth]{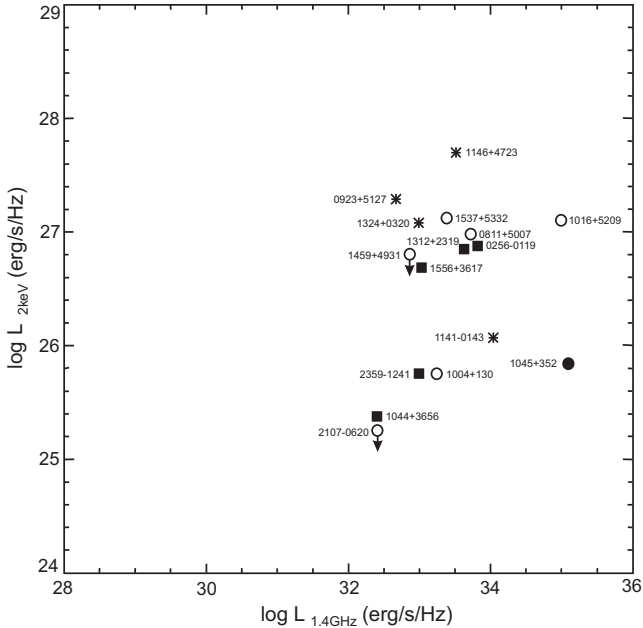}
\caption{{\footnotesize
Rest frame radio and X-ray luminosities for radio-loud BAL quasars
observed in X-rays so far:
the five BALs from \citet{broth05} sample (squares);  the first known
radio-loud BALQSO with FRII morphology 1016+5209 from \citet{schaer06};
large scale hybrid object 1004+130 from \citet{mill06}; four BALs from
\citet{wang08}, two of them have only upper limit; 1045+352 (this paper);
four radio-loud BALs from \citet{gius08}(stars).
}}
\label{Xray_plot}
\end{figure}

\citet{broth05} made  {\it Chandra} X-ray observations of five radio-loud 
BAL quasars and noticed that they are X-ray weak relative to similar
non-BAL quasars, but brighter than radio-quiet quasars
(Fig.\ref{Xray_plot}, squares). They interpreted this X-ray weakness
as a result of large intrinsic absorption columns. The above
suggestions have been confirmed by the X-ray observations of another
two radio-loud BAL quasars
\citep{mill06,schaer06} and it seemed we could draw a separate
correlation between an X-ray and radio luminosity for radio-loud BAL
quasars.  However, recent detections of X-ray bright radio-loud BAL
quasars \citep{wang08,gius08} complicated the picture.
We have updated the Brotherton et al. radio to X-ray luminosity plot
and added all radio-loud BAL quasars observed in X-rays to date
\citep{gius08,mill06,schaer06,wang08}.
We also included our 1045+352 {\it Chandra} X-ray data
point there.  The final Fig.\ref{Xray_plot} contains 16 sources and
their properties are gathered in Table~\ref{table1}.

A wide range, 3 orders of magnitude, of the X-ray luminosities
is visible on the plot and the result is similar to that
obtained for radio-quiet BAL quasars by
\citet{gius08} and \citet{shen08}. 
These authors concluded that the large spread in the $L_{X}$ values is
because of the differences in column densities and velocities of the
quasar outflows which change with the viewing angle.  It has to be put
in mind, however, that in the case of X-ray properties of radio-loud
BAL quasars we still have a very poor statistics and more observations
are needed.

We can measure the orientation angle of radio-loud BALQSOs using radio
morphology. This has been done for some of the sources in the sample
and the results are given in Table~\ref{table1}. There is a
suggestion, based on the radio brightness temperature estimation, that
four sources have polar outflows \citep{wang08}. For the other three
sources with resolved radio structures at 
5\,GHz (two large scale ones 1004+130 and 1016+5209, and the one CSS object,
1045+352), the angles between the jet axis and the line of sight were
estimated \citep{gregg00,mill06,kun07} using the core radio-to-optical
luminosity ratio defined by \citet{wills95}.
All three have a radio core and one-sided jet suggesting asymmetry,
but only in the case of 1045+352 the estimated viewing angle implies
beaming.
Each of these three sources is located in a very different part of the
parameter space in Fig.~\ref{Xray_plot}. The beamed 1045+352 with
inclination angle $<30^{\rm o}$ is more than an order of magnitude
more luminous in radio than the unbeamed 1004+130 observed at $\sim
45^{\rm o}$ and both of them have a very similar X-ray luminosity. The
unbeamed 1016+5209 at $>40^{\rm o}$ is an order of magnitude more
luminous in X-rays than 1045+352, but both have a very similar radio
luminosity. If the position of 1016+5209 in Fig.~\ref{Xray_plot} 
indicates the amount of absorption
then this source should be less absorbed than 1045+352. On the other
hand its inclination angle is larger, so the absorption column should
be larger than the one in 1045+352.  This seems in conflict with the
orientation model.
The above case is complicated by one more thing: the nature of CSS
source 1045+352 is different from that of the two large scale objects. It seems that
the jet in 1045+352 is strong and the probable counter-jet is very weak
\citep{kunert09}, but we suspect that there are environmental effects influencing the
jet's direction and emission which are difficult to estimate. We treat the
value of the viewing angle as a rough estimation.

\citet{white07} have shown that
BALs are not found among the brightest radio-emitting quasars,
although below 2\,mJy they are systematically brighter than non-BAL 
objects, with the greatest disparity arising in LoBALs. This may
indicate that we are looking closer to the jet axis in quasars with
BALs. However the Doppler factor for these sources is rather small
(like in the case of 1045+352), and the beaming effect causes
inappreciable shift of the sources to the left on the plot
(Fig.~\ref{Xray_plot}).

\subsection{Origin of X-ray Emission in BAL Quasars}

As it has been suggested by some authors the X-ray emission of some
BAL quasars can be efficiently absorbed by dense gas close to the BLR
region. However, in the case of radio-loud BAL quasars the observed
X-ray emission might originate from relativistic jets at the further
distance from the core, where there is no shielding gas \citep{broth05,
wang08, mill09}. As has been shown in the case of 1004+130 \citep{mill06}, the
X-ray emission of the radio structure is limited to its very inner
part (the radio core and the part of the jet closest to radio
core). This is also what we have assumed in our model of X-ray
emission of 1045+352. We have shown on the example of 1045+352, that
the X-ray emission of the potential young radio jets in BAL quasars
can contribute to the observed X-ray emission of the whole source
(Fig.~\ref{corona}). It, however, depends on the angle between the jet
axis and the line of sight. It is not obvious how the X-ray absorbing
gas close/in the BLR region is spaced and if this is a common scheme
for all BAL quasars. It could be that more than one scattering path is
present there \citep{lamy}. If so the radio-jet orientation can give
us only an estimation of the X-ray contribution coming from the radio
jets, and not necessarily give us the probability of the BAL
visibility and the total X-ray emission of the source.
In the case of 1045+352 we suspect that the X-ray emission of the corona is
strongly absorbed and what we observe is mostly the X-ray emission from the
relativistic jet which seems to be very strong \citep{kun07}.

\subsection{Radio Properties and Evolution of BAL Quasars}

1045+352 quasar is a CSS radio source with the size of 4.3\,kpc.
Currently, there are only 17 compact radio-loud BAL quasars observed
with VLBI in the literature
\citep{jiang03,kun07,liu08,monte09}. About half of them have still
unresolved radio structures even in the high resolution observations,
the other have core-jet structures indicating some re-orientation or
very complex morphology, suggesting a strong interaction with the
surrounding medium \citep{kun07}. All of them are potentially smaller
than their host galaxies. The analysis of the spectral shape,
variability and polarization properties of some of them shows that
they are similar to CSS and GPS objects, and are not oriented along a
particular line of sight \citep{monte09}. However, it may suggest that 
BALs are associated with an early stage of the radio source evolution
and as has been pointed out by \citet{gregg00,gregg06} the large scale
radio-loud BALQSOs could be objects with restarted activity.

The X-ray properties of GPS/CSS objects and its correlation with the radio
data have been recently discussed \citep{siemi2008,tengstrand}. According to
the authors the correlation between the X-ray and radio luminosities can be
different for different group of radio sources. GPS quasars are not
absorbed, in contrast to GPS galaxies, which show high X-ray column
densities. The radio core luminosities for the GPS galaxies are higher than
for FR\,II sources but their X-ray luminosities are comparable. This can
indicate that only the X-ray emission of the very inner part of the radio 
structure may contribute to the X-ray emission of the whole source.
Another suggestion is that making X-ray to radio luminosity correlation 
we should treat large scale and compact BAL quasars separately.  

It has been also
mentioned by several authors \citep{rb97, kun06, siemi2008} about the
possibility
of the intermittent activity in young AGNs, which can be caused by the
accretion disk instabilities \citep{hat01, jan04, czer09}. This can lead to
the scenario, where we have phases in quasar lifetime dominated
alternatingly by X-ray emission from the base of the radio jet or the
accretion disk corona.

\section{Summary}
\label{summary}

We presented {\it Chandra} X-ray observations and describe spectral
properties of a bright radio-loud BAL quasar 1045+352. We suggest that
the observed X-ray emission could be due to both the disk corona and
the relativistic jet. 
The X-ray emission due to the jet SSC emission is quite significant
and may dominate the X-ray energy range. On the other hand IC
scattering emission due to the radiation field produced by dust and
stars in the center of the galaxy is negligible in this source. 
Deeper X-ray observations of 1045+352 will be very useful to constrain
possible model's solutions.

We conclude that there is no correlation between the radio and X-ray
luminosities present in a sample of radio-loud BAL quasars observed in
X-rays to date.
A wide range of X-ray emission values indicates a more complicated
relationship between X-ray emission, radio emission and orientation based on
the radio morphology. 
There is also still an open question about the
evolutionary status of radio-loud BAL quasars. 
From the radio observations, we may conclude
that most of them are compact and probably young radio sources. It is
possible then that in the case of radio-loud BAL quasars, BALs are visible
until the radio jets escape the host galaxy.

\acknowledgements

We thank Bo\.zena Czerny for helpful discussions. This research is
funded in part by NASA contract NAS8-39073 and {\it Chandra} Award Number 
GO8-9115X issued by the
Chandra X-Ray Observatory Center, which is operated by the Smithsonian
Astrophysical Observatory. 
This work was supported by the Polish Ministry of Science and
Higher Education under grant N N203 303635.

\clearpage

\begin{landscape}
\begin{table*}
{\normalsize
\caption[]{Basic parameters of 16 radio-loud BAL quasars. \label{table1}}
\begin{tabular}{@{}l c c c r c c r c l c c c c c@{}}
\tableline
\tableline
\multicolumn{1}{l}{Source} &
\multicolumn{1}{c}{RA} &
\multicolumn{1}{c}{Dec} &
\multicolumn{1}{c}{\it z}&
\multicolumn{1}{c}{$F_{1.4}$}&
\multicolumn{1}{c}{log$L_{1.4}$}&
\multicolumn{1}{c}{$\alpha_{1.4}^{4.85}$}&
\multicolumn{1}{c}{$F_{\rm X}$}&
\multicolumn{1}{c}{log$L_{\rm X}$}&
\multicolumn{1}{c}{$\Gamma$}&
\multicolumn{1}{c}{$\alpha_{ox}$}&
\multicolumn{1}{c}{Type}&
\multicolumn{1}{c}{LLS}&
\multicolumn{1}{c}{Radio}&
\multicolumn{1}{c}{Angle}\\

\multicolumn{1}{l}{Name}& 
\multicolumn{1}{c}{h~m~s} & 
\multicolumn{1}{c}{$\degr$~$\arcmin$~$\arcsec$} &
\multicolumn{1}{c}{}&
\multicolumn{1}{c}{}&
\multicolumn{1}{c}{}& 
\multicolumn{1}{c}{}&
\multicolumn{1}{c}{}&
\multicolumn{1}{c}{}&
\multicolumn{1}{c}{}& 
\multicolumn{1}{c}{}&
\multicolumn{1}{c}{}&
\multicolumn{1}{c}{}&
\multicolumn{1}{c}{morph.}&
\multicolumn{1}{c}{}\\ 

\multicolumn{1}{c}{(1)}& 
\multicolumn{1}{c}{(2)}& 
\multicolumn{1}{c}{(3)}&
\multicolumn{1}{c}{(4)}&
\multicolumn{1}{c}{(5)}&
\multicolumn{1}{c}{(6)}&
\multicolumn{1}{c}{(7)}&
\multicolumn{1}{c}{(8)}&
\multicolumn{1}{c}{(9)}&
\multicolumn{1}{c}{(10)}&
\multicolumn{1}{c}{(11)}&
\multicolumn{1}{c}{(12)}&
\multicolumn{1}{c}{(13)}&
\multicolumn{1}{c}{(14)}&
\multicolumn{1}{c}{(15)}\\

\tableline
0256-0119      & 02 56 25.56&-01 19 12.10&2.490&27.56&33.88&0.5&5.04&26.86&1.70&
               1.90&Hi&$<$14&S&$-$\\
0811+5007      & 08 11 02.91& 50 07 24.57&1.838&24.93&33.55&0.5&12.31&26.95&1.71
               $\ast$&1.55&Lo&$<$14&S&P\\
0923+5127      & 09 23 45.19& 51 27 10.10&2.163&1.79&32.58&0.5&15.60&27.21&1.70&
               1.59&Hi&$<$33&S&$-$\\
1004+130       & 10 07 26.10& 12 48 56.20& 0.240&1126.00&33.26&0.8&96.50&25.78&1.50
               $\ast$&1.87&Hi&383.5&H&$\sim45^{\rm o}$\\
1016+5209      & 10 16 14.26& 52 09 15.45&2.455&176.90&34.99&1.1&96.10&27.13&1.70&
               1.06(1.19)&Lo&370.0&D& $>40^{\rm o}$\\
1044+3656      & 10 44 59.59& 36 56 05.39& 0.701&15.00&32.40&0.5&3.41&25.42&1.70&
               2.0(2.1)&FeLo&$<$13&S&$-$\\
1045+352       & 10 48 34.25& 34 57 24.99&1.604&1051.00&35.13&0.7&1.32&25.84&1.70&
               1.38(1.88)&Hi&4.3& Cj?&$<30^{\rm o}$\\
1141-0143      & 11 41 11.62&-01 43 06.70&1.266&141.10&33.98&0.6&4.01&26.09&1.70&
               2.01&Lo&176.6&D&$-$\\
1146+4723      & 11 46 36.88& 47 23 13.35&1.895&24.05&33.56&0.5&65.00&27.70&1.70& 
               1.38&Lo&$<$12&aD?&$-$\\
1312+2319      & 13 12 13.58& 23 19 58.64&1.520&44.12&33.74&0.8&15.77&26.87&1.70&
               1.40&Hi&1.3&Cj?&$-$\\
1324+0320      & 13 24 01.53& 03 20 20.60& 3.371&1.50&32.88&0.5&3.82&27.03&1.70&
               1.89&H&$<$41&S&$-$\\
1459+4931      & 14 59 26.33& 49 31 36.86&2.370&4.74&33.06&0.5&4.01&26.82&1.90&
               1.77&Lo&$<$45&S&P\\
1537+5332      & 15 37 03.94& 53 32 19.97& 2.403&9.58&33.38&0.5&16.95&27.15&1.31
               $\ast$&1.73&H&$<$9&S&P\\
1556+3517      & 15 56 33.77& 35 17 57.62&1.480&31.72&33.29&0.1&10.56&26.67&1.70&
               0.8(2.0)&FeLo&$<$9&S&$-$\\
2107-0620      & 21 07 57.67&-06 20 10.66& 0.645&19.61&32.43&0.5&2.86&25.30&1.90&
               1.93&FeLo&$<$8&S&P\\
2359-1241      & 23 59 53.63&-12 41 47.92& 0.870&39.50&33.03&0.5&4.55&25.76&1.70&
               1.8(2.2)&Lo&$<$134&S&$-$\\
\tableline
\end{tabular}

\vspace{0.5cm}
Description of the columns:
(1) source name;
(2) source right ascension (J2000);
(3) source declination (J2000);
(4) redshift;
(5) total flux density at 1.4\,GHz (mJy);
(6) log of the radio luminosity at a rest-frame of 1.4\,GHz ($\rm
erg~s^{-1}~Hz^{-1}$);
(7) spectral index between 1.4 and 4.85\,GHz;
(8) total flux density at 2\,keV ($\rm
10^{-15}~erg~s^{-1}~cm^{-2}~keV^{-1}$) calculated assuming
the Galactic absorption and photon index;
(9) log of the X-ray luminosity at a rest-frame of 2\,keV ($\rm
erg~s^{-1}~Hz^{-1}$);
(10) photon index, '$\ast$' means fitted, in other case - assumed;
(11) the optical - X-ray spectral index; quantities in parentheses are
calculated using dereddened optical flux;
(12) BALQSO subclassification, Hi - HiBAL, Lo - LoBAL, FeLo - FeLoBAL, H - BALQSO for 
which the Mg\,II spectral region is redshifted outside the SDSS window; 
(13) largest linear size ($h^{-1}~{\rm kpc}$), '$<$' means upper limit of the linear size
estimated using deconvolved component major axis angular size from FIRST or
NVSS when available or using FIRST beam (5\farcs4); 
(14) radio morphology, S - single means unresolved in available radio image, Cj - core-jet, D
- double, aD - asymmetric double, H - hybrid object (FR\,I/FR\,II);
(15) angle between the jet axis and the observer based on the radio image, P - polar
outflow.
The information about the radio-loud BAL quasars are gathered from: \citet{broth05},
\citet{gius08}, \citet{gregg00}, \citet{kun07}, \citet{mill06},
\citet{schaer06}, \citet{wang08}.

}
\end{table*}
\clearpage
\end{landscape}

\end{document}